\documentclass{JHEP}

\title{Duality and the Equivalence Principle of Quantum Mechanics}
\author{Jos\'e M. ISIDRO \\ Dipartimento di Fisica 
``G. Galilei'' -- Istituto Nazionale di Fisica Nucleare\\ Universit\`a di Padova,
Via Marzolo, 8 -- 35131 Padova, Italy
\email{isidro@pd.infn.it}}

\abstract{Following a suggestion by Vafa, we present a quantum--mechanical model 
for S--duality symmetries observed in the quantum theories of fields, strings and branes. 
Our formalism may be understood as the topological limit of Berezin's metric quantisation 
of the upper half--plane ${\bf H}$, in that the metric dependence of Berezin's method 
has been removed. Being metric--free, our prescription makes no use of global quantum numbers. 
Quantum numbers arise only locally, after the choice of a local vacuum to expand around. 
Our approach may be regarded as a manifestly non--perturbative formulation of quantum mechanics, 
in that we take no classical phase space and no Poisson brackets as a starting point. 
Position and momentum operators satisfying the Heisenberg algebra are defined 
and their spectra are analysed. We provide an explicit construction 
of the Hilbert space of states. The latter carries no representation of $SL(2,{\bf R})$,  
due to the lifting of the metric dependence. Instead, the reparametrisation
invariance of ${\bf H}$ under $SL(2,{\bf R})$ induces a natural $SL(2,{\bf R})$ 
action on the quantum--mechanical operators that implements S--duality. 
We also link our approach with the equivalence principle of quantum mechanics 
recently formulated by Faraggi--Matone.}


\preprint{DFPD00/TH/46, US-FT/14-00}

\begin{document}

\section{Introduction}\label{intro}

\subsection{Motivation}\label{motivation}

The concept of {\it duality} plays a key role in important recent developments in the 
quantum theories of fields \cite{QFTDUAL}, string duality \cite{STRINGDUALITY}, 
M--theory and branes \cite{MTHEORY}, M(atrix) theory \cite{MATRIX}, and the AdS/CFT 
correspondence \cite{ADS}. Broadly speaking, under duality one understands 
a transformation of a given field or string theory, in a certain regime of 
the variables and parameters that define it, into a physically equivalent theory 
with different variables and parameters. The theories thus mapped into 
each other may be of apparently very different nature ---{\it e.g.}, the 
duality may exchange a field theory with a string theory. Alternatively, the duality may exchange 
the strong--coupling regime of a given theory with the perturbative regime 
of its dual theory, thus making the former more tractable. This latter 
form of mapping different theories goes under the name of S--duality. 
Often, what appears to be a highly non--trivial quantum excitation in a given 
field or string theory may well turn out to be a simple perturbative correction 
from the viewpoint of a theory dual to the original one. This suggests that what constitutes 
a quantum correction may be a matter of convention: the notion of {\it classical} 
versus {\it quantum} is relative to which theory the measurement is made from. 

In view of these developments, Vafa \cite{VAFA} and other authors have suggested 
that quantum mechanics itself may need a revision if it is to accommodate, 
already from first principles, the notion of duality.

This state of affairs is reminiscent of general relativity. In fact, 
Faraggi--Matone \cite{MATONE} and Matone \cite{MMATONE} have recently developed
a very interesting formulation of quantum mechanics from an equivalence principle 
resembling that of general relativity. In this formulation, conformal symmetry plays a key role.

Conformal quantum mechanics, as initiated in \cite{ALFARO} and later supersymmetrised 
in \cite{RABINOVICI}, has also been the subject of renewed interest in connection with 
multi--black hole quantum mechanics (see \cite{DFG} for extensive references) and 
AdS spaces \cite{SOROKIN}. 

\subsection{Summary}\label{summary}

Motivated by the above considerations, in this paper we develop a quantum 
mechanics that naturally incorporates a simple form of S--duality. The 
latter will be modelled on the conformal transformation of a complex variable 
$z\rightarrow -z^{-1}$. Due to the presence of conformal symmetry, 
our formalism may also be understood as the appropriate quantum mechanics 
for the affine variables of quantum gravity
\footnote{In the context of quantum gravity the affine algebra plays a significant role
\cite{ISHAMKAKAS}.  The generator of translations is represented by an operator 
with strictly positive spectrum. A similar feature will appear in our 
formalism in section \ref{qmhft}. When generalising the 1--dimensional affine algebra 
to several dimensions \cite{AFFKLAUDER}, the generator of translations becomes a symmetric, 
positive--definite matrix degree of freedom. Such an object is well suited to describe 
the spatial part of the metric tensor. The coherent--state representation of the 1--dimensional 
affine algebra has been studied in \cite{1DIMAFF}, and it has been generalised 
to several dimensions in \cite{WATSON}. An interesting application of the Faraggi--Matone 
duality to gravity and Dirac fields has been obtained in \cite{BRASIL}.}.

The presence of conformal symmetry suggests considering a variable defined 
on Poincar\'e's upper half--plane ${\bf H}$. On the latter there exists the 
{\it holomorphic Fourier transformation} (HFT), which we intend to use as a technical tool 
for quantising an affine variable. The HFT relates a real coordinate on ${\bf R}$ 
to a complex momentum on ${\bf H}$. Alternatively, a real momentum can be HFT--transformed 
into a complex coordinate on ${\bf H}$.  Position and momentum operators satisfying 
the Heisenberg algebra will be defined, as dictated by the HFT. The  Hilbert space of states 
will be identified explicitly. It will turn out to be larger than the 
standard $L^2({\bf R})$ Hilbert space, as a consequence of the 
non--perturbative nature of our quantisation. We will explain how it 
eventually reduces to the usual $L^2({\bf R})$. The wave function on ${\bf R}$ will 
be the restriction to the boundary of a holomorphic wave function whose natural domain 
will be ${\bf H}$. However, the quantum--mechanical operator $Z$ corresponding 
to the classical variable $z\in {\bf H}$ will not be self--adjoint, 
so its physical interpretation requires some care. One can nonetheless make sense 
out of a non self--adjoint operator $Z$. This is based on the fact that $Z^2$ 
admits a self--adjoint Friedrichs extension, whose square root is now self--adjoint.

Our formalism may be understood as a certain limit of Berezin's quantisation \cite{BEREZIN}. 
The latter relies on the metric properties of classical phase space ${\cal M}$, 
whenever ${\cal M}$ is a homogeneous K\"ahler manifold. In Berezin's method, 
quantum numbers arise naturally from the metric on ${\cal M}$. The semiclassical regime 
is then identified with the regime of large quantum numbers. Our method may be regarded 
as the topological limit of Berezin's quantisation, in that the metric dependence 
has been removed. Topological gravity has in fact a long history \cite{WITTEN}. 
As a consequence of this topological nature our quantisation exhibits some added features. 
Quantum numbers are not originally present in our prescription; they appear only 
after a vacuum has been chosen, and even then they are local in nature, instead of global. 
Hence our procedure may be thought of as a manifestly non--perturbative formulation 
of quantum mechanics, in that we take no classical phase space and no Poisson brackets 
as our starting point, {\it i.e.}, we do not deform a classical theory into 
its quantum counterpart, as in deformation quantisation \cite{KONTSEVICH, CATTANEO, 
SCHLICHENMAIER, FELDER}.

On the upper half--plane ${\bf H}$ there is an isometric action of the group $SL(2,{\bf R})$.
Berezin's  metric method, applied to ${\bf H}$, yields a Hilbert space of states ${\cal H}$ 
that provides a representation space for $SL(2,{\bf R})$. However,  our approach makes
no use of the metric  properties of ${\bf H}$. Correspondingly, we have no representation of
$SL(2,{\bf R})$ as a Hilbert space of states. Also in this sense our quantisation is topological, 
as opposed to Berezin's metric approach,  and it bears some resemblance with topological field 
theory \cite{LOZANO}.

\subsection{Outline}\label{outline}

This paper is organised as follows. Section \ref{berezin} sets the scene by giving 
a quick review of Berezin's quantisation, starting from the metric on certain classical 
phase spaces.  The HFT is presented in technical detail in section \ref{hft}. 
Section \ref{qmhft} develops a quantum mechanics based on the HFT. Special emphasis is placed 
on a technical analysis of the spectral properties of operators.  Section \ref{discussion} 
is devoted to a physical interpretation of our formalism. We discuss why the non--isospectrality 
of the HFT allows for non--trivial dualities that are necessarily absent 
in the context of Schr\"odinger pairs, as in standard quantum mechanics.
We also explain the physical meaning of the non self--adjoint operator $Z$, 
the choice of a vacuum and the breaking of $SL(2,{\bf R})$ to the affine group
of quantum gravity, as well as the topological character of this quantum mechanics. 
Using an $SL(2,{\bf R})$ action on the operators, we exhibit how to implement an 
S--duality between strong quantum effects and semiclassical corrections in our framework. 
In section \ref{final} we comment on the link between our work and the new approach 
to quantum mechanics from the equivalence principle of Faraggi--Matone \cite{MATONE, MMATONE}.
One might regard the HFT merely as a technical tool that allows for a simple 
quantum--mechanical implementation of S--duality. However, looking beyond,
in section \ref{outlook} we also propose a possible starting point 
for an explicitly non--perturbative formulation of quantum mechanics in 
the sense claimed by Vafa \cite{VAFA}.

\section{Berezin's Quantisation}\label{berezin}

For later use we will first sketch the construction of the Hilbert space
of states ${\cal H}$ from the metric on some relevant homogeneous K\"ahler manifolds
\cite{BEREZIN, KLAUDER}.

\subsection{The Lobachevsky Plane}\label{lobachevsky}

Consider the Lobachevsky plane modelled as the unit disc
$D=\{z\in {\bf C}:|z|<1\}$. From the K\"ahler potential $K_{D}(z,\bar z)=-{\rm log}\,(1-|z|^2)$ 
one derives an integration measure ${\rm d}\mu(z,\bar z)=
(2\pi {\rm i})^{-1}{\rm d} z\wedge {\rm d}\bar z/ (1-|z|^2)^2$.  
With respect to the scalar product
\begin{equation}
\langle\psi_1|\psi_2\rangle =\Big({1\over \hbar} -1\Big)\int_{D}{\rm d}\mu(z,\bar
z)\,(1-|z|^2)^{1/\hbar}\,{\overline\psi_1(z)}\,\psi_2(z),
\label{lobscalar}
\end{equation} 
the space ${\cal F}_{\hbar}(D)$ of analytic functions on $D$ with finite norm defines a  Hilbert
space of states ${\cal H}$. Setting $k-1=(2\hbar)^{-1}$,  ${\cal F}_{\hbar}(D)$ becomes the
representation space for the discrete series of the group $SU(1,1)$ of isometries of $D$, {\it
i.e.}, the space of weight-$k$ modular forms. Large values of $k$ correspond to the semiclassical
limit of this quantum mechanics. Coherent states $|w\rangle$ are parametrised by points $w$ in the
quotient space $SU(1,1)/U(1)$.

\subsection{Complex Homogeneous K\"ahler Manifolds}\label{complex}

Let $z^j$, $\bar z^k$, $j, k= 1,\ldots, n$, be  local coordinates on a
complex homogenous K\"ahler manifold ${\cal M}$, and let $K_{\cal M}(z^j,\bar z^k)$ be a K\"ahler
potential for the metric ${\rm d} s^2=g_{j\bar k}\,{\rm d} z^j{\rm d}\bar z^k$. The K\"ahler form
$\omega=g_{j\bar k}\,{\rm d} z^j\wedge {\rm d}\bar z^k$ gives rise to an integration measure
${\rm d}\mu(z,\bar z)$, 
\begin{equation} 
{\rm d}\mu(z,\bar z)=\omega^n={\rm det}\,(g_{j\bar k})\,\prod_{l=1}^n{{\rm d} z^l\wedge {\rm d}\bar
z^l\over 2\pi {\rm i}}.
\label{bermeasure}
\end{equation} 
The Hilbert space of states ${\cal H}$ is the space ${\cal F}_{\hbar}({\cal M})$ of analytic
functions on
${\cal M}$ with finite norm, the scalar product being
\begin{equation}
\langle\psi_1|\psi_2\rangle =c(\hbar)\,\int_{\cal M}{\rm d}\mu(z,\bar z)\,{\rm
exp}(-\hbar^{-1}K_{\cal M}(z,\bar z))\,{\overline
\psi_1(z)}\psi_2(z),
\label{berscalar}
\end{equation}
and $c(\hbar)$ a normalisation factor.  Let  $G$ denote the Lie group of motions of ${\cal M}$, 
and assume $K_{\cal M}(z,\bar z)$ is invariant under $G$. Setting $\hbar=k^{-1}$, the family
of Hilbert spaces ${\cal F}_{\hbar}({\cal M})$ provides a discrete series of projectively unitary
representations of $G$. The homogeneity of ${\cal M}$ is used  to prove that the correspondence
principle is satisfied in the limit $k\to\infty$. Furthermore, let $G'\subset G$ be a maximal
isotropy subgroup of the vacuum state $|0\rangle $. Then coherent states $|\zeta\rangle $ are
parametrised by points
$\zeta$ in the coset space $G/G'$.

\section{The Holomorphic Fourier Transformation (HFT)}\label{hft}

By analogy with section \ref{berezin}, we need a space of analytic functions 
on the upper half--plane ${\bf H}$ as our Hilbert space of states ${\cal H}$. 
A key observation is that the holomorphic Fourier transformation, summarised below, 
provides such a space in a natural way \cite{YOSIDA}. 

Let  $F_{\psi}\in L^2(0,\infty)$. For
$z=x+{\rm i} \,y\in {\bf H}$, the function $\psi$ defined as
\begin{equation}
\psi(z)={1\over\sqrt {2\pi}}\int_0^{\infty}{\rm d} t \,F_{\psi}(t) \, {\rm e}^{{\rm i} tz},
\label{hfourier}
\end{equation}
the integral understood in the sense of Lebesgue, is holomorphic on ${\bf H}$. 
Its restrictions to horizontal straight lines $y={\rm const}>0$ in ${\bf H}$ 
are a bounded set in $L^2({\bf R})$.

Conversely, let $\psi$ be holomorphic on ${\bf H}$, and assume that 
\begin{equation}
{\rm sup}_{ 0<y<\infty}\, \int_{-\infty}^{\infty} {\rm d} x \,|\psi(x+{\rm i} y)|^2=C<\infty.
\label{sup}
\end{equation}
Then the function $F_{\psi}$ defined by      
\begin{equation} 
F_{\psi}(t)={1\over \sqrt {2\pi}}\int_{-\infty}^{\infty}{\rm d} z \,\psi (z)\, {\rm e}^{-{\rm i}
tz},
\label{ihfourier}
\end{equation}
the integration being along any horizontal straight line $y= {\rm const}>0$ in ${\bf H}$,
satisfies the following basic properties. $F_{\psi}(t)$ is  independent of the particular
horizontal line $y= {\rm const}>0$ chosen. Moreover,  $F_{\psi}\in L^2(0,\infty)$, and 
for any $z\in {\bf H}$, equation (\ref{hfourier}) holds, with 
\begin{equation}
\int_{0}^{\infty}{\rm d} t \,|F_{\psi}(t)|^2 = C.
\label{norma}
\end{equation}
We call $F_{\psi}$ the holomorphic Fourier transform of $\psi$.

Some features of the HFT on ${\bf H}$ are worth mentioning. Let $\Omega({\bf H})$ 
denote the space of all holomorphic functions on
${\bf H}$, and let $\Omega_0({\bf H})$ denote the proper subspace of all
$\psi\in\Omega({\bf H})$ such that the supremum $C$ introduced  in (\ref{sup}) is finite. 
Then $C$ defines a squared norm $||\psi||^2$ on $\Omega_0({\bf H})$. The subspace  
$\Omega_0({\bf H})$ is complete with respect to this norm. This norm is Hilbert, {\it i.e.}, 
it verifies the parallelogram identity. Hence the scalar product $\langle\varphi|\psi\rangle$  
defined on $\Omega_0({\bf H})$ through
\begin{equation} 
4\langle \varphi|\psi\rangle=
||\psi+\varphi||^2-||\psi-\varphi||^2+{\rm i}\,||\psi+{\rm
i}\,\varphi||^2-{\rm i}\,||\psi-{\rm i}\,\varphi||^2
\label{polar}
\end{equation}
turns the complete normed space $\Omega_0({\bf H})$ into a Hilbert space with respect to the
scalar product (\ref{polar}). In fact,  via the HFT, the subspace $\Omega_0({\bf H})$ is
isometrically isomorphic to the Hilbert space $L^2(0,\infty)$.

\section{Quantum Mechanics from the HFT}\label{qmhft}

\subsection{Hilbert Space of States}\label{hilbert}

Section \ref{hft} allows us to identify the Hilbert space of states ${\cal H}$ of our quantum
mechanics. In the representation in which the wave function is $F_{\psi}(t)$ we have
${\cal H}=L^2(0,\infty)$, while in its HFT--transformed representation $\psi(z)$ we have
${\cal H}=\Omega_0({\bf H})$. After the choice of a vacuum in section 
\ref{principle} and the introduction of the boundary wave function in 
section \ref{boundary}, we will see the emergence of the usual Hilbert space 
$L^2({\bf R})$. 

For definiteness, we choose the complex variable $z\in {\bf H}$ to stand for the momentum $p$, 
with the real variable $t\in (0,\infty)$ standing for the coordinate $q$. Then the HFT reads
\begin{eqnarray}
\psi(p)&=&{1\over \sqrt{2\pi\hbar}}\int_0^{\infty}{\rm d} q\, F_{\psi}(q)\,{\rm e} ^{{{\rm i}\over
\hbar}qp}\cr F_{\psi}(q)&=&{1\over \sqrt{2\pi\hbar}}\int_{-\infty}^{\infty}{\rm d} p\, \psi(p)\, 
{\rm e}^{-{ {\rm i}\over \hbar}qp}.
\label{qreal}
\end{eqnarray}

\subsection{Operators}\label{operators}

In coordinate representation, we define position and momentum operators $Q$ and $P$:
\begin{equation}
(QF_{\psi})(q)=q\,F_{\psi}(q), \qquad (PF_{\psi})(q)={\rm i}\hbar\,
{{\rm d} F_{\psi}\over {\rm d} q}.
\label{pdef}
\end{equation}
Equation (\ref{qreal}) implies that their momentum representation is
\begin{equation}
(Q\psi)(p)=-{\rm i}\hbar\, {{\rm d}\psi\over {\rm d} p},\qquad (P\psi)(p)=p\,\psi(p).
\label{zdef}
\end{equation}
Irrespective of the representation chosen we have the Heisenberg algebra 
\begin{equation}
[P, Q]= {\rm i}\hbar\,{\bf 1}.
\label{hei}
\end{equation}
On the domain
\begin{equation} 
D(Q)=\{F_{\psi}\in L^2(0,\infty):\; \int _0^{\infty}{\rm d} q\, q^2 
|F_{\psi}(q)|^2<\infty\},
\label{domq}
\end{equation}
which is dense in ${\cal H}$, the operator $Q$ is symmetric,
\begin{equation}
\langle F_{\psi}|Q|F_{\varphi}\rangle^*=\langle F_{\varphi}|Q|F_{\psi}\rangle.
\label{barp}
\end{equation} 
A closed, symmetric, densely defined operator admits a self--adjoint extension if and only if 
its defect indices $d_{\pm}$ are equal. Moreover, such an operator is essentially self--adjoint
if and only if its defect indices are both zero \cite{YOSIDA}. The operator $Q$ 
turns out to be essentially self--adjoint, with  point, residual and continuous spectra given by
\begin{equation}
\sigma_p(Q)=\phi, \qquad \sigma_r(Q)=\phi, \qquad \sigma_c(Q)=[0,\infty).
\label{pspectra}
\end{equation}

The properties of the momentum operator $P$ are subtler. One finds 
\begin{equation} 
{\langle F_{\psi}|P| F_{\varphi}\rangle }^*={\rm i}\hbar\, F_{\psi}(0)F_{\varphi}^*(0) +
\langle F_{\varphi}|P| F_{\psi}\rangle,
\label{barzop}
\end{equation} 
so $P$ is symmetric on the domain of absolutely continuous functions
\begin{equation}
D(P)=\{ F_{\psi}\in L^2(0,\infty): 
F_{\psi}\; {\rm abs.}\; {\rm cont.}, 
\int_0^{\infty}{\rm d} q\,|{{\rm d} F_{\psi}\over {\rm d} q}|^2 <\infty,\; 
F_{\psi}(0)=0\}.
\label{domz}
\end{equation}
The adjoint $P^{\dagger}$ also acts as ${\rm i}\,\hbar\,{\rm d}/{\rm d} q$, with a
domain $D(P^{\dagger})$
\begin{equation} 
D(P^{\dagger})=\{ F_{\psi}\in L^2(0,\infty): F_{\psi}\; {\rm abs.}\; {\rm cont.}, 
\int_0^{\infty}{\rm d} q\,|{{\rm d} F_{\psi}\over {\rm d} q}|^2 
<\infty\},
\label{domzdag}
\end{equation}
where the boundary condition $F_{\psi}(0)=0$ has been lifted. On the space $L^2(0,\infty)$ 
we have $d_{+}(P)=0$, $d_{-}(P)=1$. We conclude that  $P$ admits no self--adjoint extension. 
Its point, residual and continuous spectra are
\begin{equation}
\sigma_p(P)=\phi, \qquad \sigma_r(P)={\bf H}\cup {\bf R}, \qquad \sigma_c(P)=\phi.
\label{zspectra}
\end{equation}

The domain $D(P)$ is strictly contained in $D(P^{\dagger})$. This implies that the operators
$P_x=(P+P^{\dagger})/2$ and $P_y=(P-P^{\dagger})/2{\rm i}$ which one would naively construct
out of $P$  are ill defined.  There is no way to define self--adjoint operators $P_x$ and $P_y$ 
corresponding to the classical momenta $p_x$ and $p_y$. This is compatible with the fact that,
the defect indices of $P$ being unequal, $P$ does not commute with any complex conjugation on
${\cal H}$ \cite{YOSIDA}. However, we will see presently that one can 
make perfectly good sense of a quantum mechanics whose momentum operator $P$ 
admits no self--adjoint extension.
\footnote{Standard quantum mechanics associates a self--adjoint operator with each observable. 
A more refined mathematical description has been proposed in terms of positive--operator--valued  
measures \cite{POVM, EGUS}. These ideas have been applied to  certain problems that escape the usual
quantum--mechanical approach, such as the time observable, which cannot be described by a
self--adjoint operator \cite{PAULI} (see however \cite{GALAPON}). In \cite{JAEKEL} 
it has been shown that the operators describing the coordinates of an event can be 
determined by imposing covariance under the conformal group. Again, these operators cannot 
be self--adjoint \cite{WIGHTMANN}, but they have been successfully treated in \cite{TOLLER} 
by means of conformally covariant, positive--operator--valued measures. 
I thank I. Egusquiza for an interesting discussion on the subject.}
We defer issues like measurements of $P$ 
and Heisenberg's uncertainty principle until section \ref{principle}. 
Quadratic terms in $P$ are technically simpler, and will be dealt with first.

With our choice of domain $D(P)$, which makes $P$ symmetric, $P^2$ is also symmetric.
One proves that $d_-(P^2)=1=d_+(P^2)$. Hence $P^2$, although not essentially self--adjoint, 
admits a self--adjoint extension. A popular choice is the Friedrichs extension \cite{YOSIDA}. 
Given an operator $A$, this extension is characterised by a boundedness condition
\begin{equation}
\langle \varphi|A|\varphi\rangle \geq -\alpha \,||\varphi||^2 \qquad \forall \varphi\in D(A)
\label{friedbound}
\end{equation}
for a certain $\alpha\geq 0$. Now the operator $P^2$ admits a Friedrichs extension $P^2_F$ 
with a lower bound
$\alpha=0$:
\begin{equation}
\langle F_{\varphi}|P^2_F|F_{\varphi}\rangle \geq 0, \qquad \forall F_{\varphi}\in D(P^2_F).
\label{pbound}
\end{equation}
The point, residual and continuous spectra of this extension are
\begin{equation}
\sigma_p(P^2_F)=\phi,\qquad \sigma_r(P^2_F)=\phi, \qquad \sigma_c(P^2_F)=[0,\infty).
\label{ptwospectra}
\end{equation}

Now the crucial point is that the square root of the Friedrichs extension 
allows us to define a self--adjoint momentum operator. 
Let us define the new operator $P_{\sqrt{}}$ 
\begin{equation}  
P_{\sqrt{}}=\sqrt{P^2_F}.
\label{square}
\end{equation}  
$P_{\sqrt{}}$ is self--adjoint, with a domain $D(P_{\sqrt{}})$ uniquely
determined by the spectral decomposition of $P$ \cite{YOSIDA}. 
The point, residual and continuos spectra of $P_{\sqrt{}}$ are
\begin{equation}
\sigma_p(P_{\sqrt{}})=\phi,\qquad \sigma_r(P_{\sqrt{}})=\phi,\qquad
\sigma_c(P_{\sqrt{}})=[0,\infty).
\label{spectrappm}
\end{equation}
We observe that taking the Friedrichs extension does not commute with the square root. 
The operator $P_{\sqrt{}}$ enjoys the properties of being linear in $p$ 
and having the right commutator (\ref{hei}) with the position operator.  

\subsection{$SL(2,{\bf R})$--Transformation of the Operators}\label{transformation}

We can reparametrise the coordinate $z\in {\bf H}$ by means of a M\"obius transformation
$z\rightarrow \tilde z=(az+b)(cz+d)^{-1}$, with $ad-bc=1$.  We now consider the HFT written as
\begin{eqnarray}
\tilde\psi(\tilde p)&=&{1\over \sqrt{2\pi\hbar}}
\int_0^{\infty}{\rm d} \tilde q\, \tilde F_{\tilde\psi}(\tilde q)\,
{\rm e}^{{{\rm i}\over\hbar}\tilde q\tilde p}\cr 
\tilde F_{\tilde\psi}(\tilde q)&=&{1\over \sqrt{2\pi\hbar}}
\int_{-\infty}^{\infty}{\rm d} \tilde p\,\tilde\psi(\tilde p)\,{\rm e}^{-{ {\rm i}\over\hbar}
\tilde q\tilde p},
\label{tildeqreal}
\end{eqnarray}
where $\tilde q\in (0,\infty)$ is the variable dual to $\tilde p$ under (\ref{tildeqreal}). 
One can define coordinate and momentum operators $\tilde Q$ and $\tilde P$ satisfying 
the Heisenberg algebra (\ref{hei}). Hence this is a canonical transformation from $(q,p)$ 
to $(\tilde q, \tilde p)$. The transformed operators $\tilde Q$ and $\tilde P$ 
have the same spectra as before.

\section{Physical Discussion}\label{discussion}

\subsection{Schr\"odinger Pairs vs. the Non--Isospectrality of the HFT}\label{pairs}

The standard Fourier transformation maps (a subspace of) $L^2({\bf R})$ into (a subspace of)
$L^2({\bf R})$. It is also an isospectral transformation between self--adjoint operators. 
In the context of the standard Fourier transformation on $L^2({\bf R})$, coordinate and 
momentum are sometimes referred to as a {\it Schr\"odinger pair}.

On the contrary, the HFT is not an isospectral transformation: $Q$ and $P$ do not have
identical spectra. Furthermore, the very choice of the dynamical variable to be represented by
complex variable $z$ of the HFT is a non--trivial choice in itself. These properties allow for
non--trivial dualities that are necessarily absent in the context of Schr\"odinger pairs. 

\subsection{The Choice of a Vacuum}\label{principle}

The difficulties due to the fact that one of the two canonical operators $(Q,P)$ admits no
self--adjoint extension can be overcome by {\it the choice of a vacuum} to expand around.
Under the latter we understand the choice of either $z$ or $\tilde z=-z^{-1}$ as the classical 
coordinate on ${\bf H}$ to be quantised into the operator $Z$ or $\tilde Z=-Z^{-1}$. 
After the choice of a vacuum, the $SL(2,{\bf R})$ symmetry is reduced to translations 
and dilatations, leaving the affine group only.

For definiteness, let us choose the vacuum corresponding to the classical variable $z$, 
in the picture in which the quantum operator $Z$ is the momentum $P$. 
Then the construction of section \ref{qmhft} leads to a pair of self--adjoint operators
$(Q,P_{\sqrt{}})$. They are almost canonically conjugate in the sense that, while satisfying
the Heisenberg algebra (\ref{hei}), the exchange between coordinate and momentum is not performed
directly at the level of $(Q,P_{\sqrt{}})$ by means of the usual Fourier transformation. 
Rather, $(Q,P_{\sqrt{}})$ have to be lifted back to their HFT ancestors $(Q,P)$, 
in order to exchange them. Apart from this technicality, the operators $(Q,P_{\sqrt{}})$ 
meet the usual quantum--mechanical requirements concerning the measurement process and 
the Heisenberg uncertainty principle. The vacuum $|0_z\rangle$, and the corresponding 
local quantum numbers $n_z$ obtained upon expansion around it, will certainly differ from 
the vacuum $|0_{\tilde z}\rangle$ and the quantum numbers $n_{\tilde z}$ obtained from 
the classical variable $\tilde z=-z^{-1}$. So this choice of a vacuum is {\it local in nature}, 
in that it is linked to a specific choice of coordinates. The coherent states constructed around 
$|0_z\rangle$ are not coherent from the viewpoint of $|0_{\tilde z}\rangle$. 
We conclude that this quantum mechanics does not allow for globally defined coherent states 
such as those of section \ref{berezin}. 

This choice of a vacuum is reminiscent of M--theory and the (perturbatively) different
string theories it unifies \cite{STRINGDUALITY, MTHEORY}. The eleventh dimension of
M--theory, as opposed to the ten critical dimensions of the type IIA string, appears 
in the passage to the strong coupling limit. In so doing one succeeds in incorporating 
the known dualities between different perturbative strings. In our context, 
the HFT canonically relates the two real dimensions of the upper half--plane ${\bf H}$ 
to the one real dimension of the real axis ${\bf R}$.  The extra dimension present 
in the HFT disappears once a vacuum has been chosen, through the self--adjoint 
operator $P_{\sqrt{}}$.

\subsection{The Boundary Wave Function}\label{boundary}

After a vacuum has been chosen, the connection with standard quantum mechanics 
can be made more explicit by exhibiting the usual Hilbert space $L^2({\bf R})$ 
emerge from our approach as follows. Let us consider the picture (dual to that of section 
\ref{qmhft}) in which the complex variable $z$ is the coordinate $q$. 
For a holomorphic wave function $\psi(q)=\psi(x+{\rm i} y)$ satisfying condition
(\ref{sup}), a boundary wave function $\psi_b(x)\in L^2({\bf R})$ exists such that
\cite{YOSIDA}
\begin{equation} 
{\rm lim}_{y\to 0}\int_{-\infty}^{\infty}{\rm d} x\, |\psi(x+{\rm i} y)-\psi_b(x)|^2=0.
\label{lim}
\end{equation} 
So while the requirement of $L^2$--integrability of standard quantum mechanics is maintained, 
the HFT extends the wave function $\psi_b(x)\in L^2({\bf R}) $ of a particle on the boundary 
of ${\bf H}$ to a holomorphic $\psi(q)$ defined on the entire upper half--plane. 

\subsection{Quantum Numbers vs. a Topological Quantum Mechanics}\label{quantumnumbers}

Berezin's quantisation relied heavily on the metric properties of
classical phase space. The semiclassical limit could be defined as the regime of large quantum
numbers. The very existence of quantum numbers was a consequence of the metric structure.

On the contrary, the quantum mechanics developed here is completely independent of the metric
properties of the upper half--plane ${\bf H}$. Quantisation in terms of the 
HFT is {\it topological}, in that it does not know about the Poincar\'e metric 
${\rm d} s^2=({\rm d} x^2+{\rm d} y^2)/y^2$. Indeed, the absence of a metric prevents us 
from writing an integration measure along the lines of section \ref{berezin}. 
The supremum in equation (\ref{sup}) reflects this fact. Along any 
horizontal line $y={\rm const}>0$ one effectively observes a constant Euclidean metric; 
in order to detect the negative curvature of the Poincar\'e metric one needs to displace along $y$.  
The HFT correctly captures this property. Furthermore, a ``thickening" of the real line ${\bf R}$ 
to the the upper half--plane ${\bf H}$ should not feel the presence of the Poincar\'e metric on 
${\bf H}$, if it is to describe quantum mechanics on ${\bf R}$. This is compatible 
with the interpretation of the wave function given in section \ref{boundary}.

We therefore have a quantum mechanics that is free of {\it global} quantum numbers.  
The latter appear only after the choice of a {\it local} vacuum. The logic could be
summarised as follows: 
\begin{itemize}
\item[1.]{the fact that this quantum mechanics is topological implies the  absence of
a metric;}
\item[2.]{the absence of a metric implies the absence of global quantum numbers;}
\item[3.]{the absence of
global quantum numbers implies the impossibility of globally defining a semiclassical regime.
The latter exists only locally.}
\end{itemize}
Actually, as our starting point we have no classical phase space at all, 
and no Poisson brackets to quantise into commutators. This may be regarded 
as a manifestly non--perturbative formulation of quantum mechanics, as 
required in \cite{VAFA}. The sections that follow elaborate on this point further. 

\subsection{Classical vs. Quantum}\label{conjectures}

Next we state a proposal to accommodate a simple form of S--duality into our framework.
To be concrete, we assume that the required duality is $SL(2,{\bf R})$. In fact this group (or
subgroups thereof) is ubiquitous in field and string duality. From the HFT we have developed a
quantum mechanics that is conceptually as close as possible to the standard one, while at the same
time  incorporating the desired duality. By this we do not mean having a representation of
$SL(2,{\bf R})$ as the Hilbert space of states. In fact Berezin's quantisation does precisely that
\cite{BEREZIN}. Rather, we have implemented a particularly relevant $SL(2,{\bf R})$ transformation,
the inversion $z\rightarrow\tilde z=-z^{-1}$, on the quantum operator $Z$ corresponding to the
classical variable $z$.  If $Z$ is taken to represent the momentum $P$, the effect is that of
transforming Planck's constant as $\hbar \rightarrow -\hbar^{-1}$. This can be interpreted as an
exchange of the semiclassical with the strong quantum regime. In this context, $\hbar$ is best
thought of as a dimensionless deformation parameter, as in section \ref{berezin} 
and in deformation quantisation \cite{KONTSEVICH, CATTANEO, SCHLICHENMAIER, FELDER}. 
This duality symmetry is not implemented in ordinary quantum mechanics. 

The quantum mechanics based on the HFT naturally incorporates this duality under a single theory.
Different limits of the latter yield different regimes. Let us start from the classical variable
$z\in {\bf H}$ and choose the corresponding non self--adjoint quantum operator $Z$ to be the
momentum $P$. We can compute quantum effects to $O(\hbar)$, which one would call semiclassical 
{\it from the viewpoint of the quantum theory corresponding to the classical variable $z$}. 
Strong quantum effects, that will be  of $O(-\hbar^{-1})$  {\it from the viewpoint of the original
theory}, will appear to be simple semiclassical corrections of $O(\hbar)$ {\it from the viewpoint
of the dual quantum theory corresponding to the classical variable $\tilde z=-z^{-1}$}.

\section{Relation with the Equivalence Principle of Faraggi--Matone}\label{final}

In \cite{MATONE} an entirely new presentation of quantum mechanics has been given, 
starting from an equivalence principle that is close in spirit to that of general 
relativity. In plain words, the philosophy underlying this approach could be summarised 
as follows. The classical Hamilton--Jacobi technique is based on transforming an arbitrary 
dynamical system, by means of coordinate changes, into a freely evolving system 
subject to no interactions. The requirement that this equivalence also hold 
in the case in which the conjugate variables are considered as dependent 
leads to a quantum analogue of the Hamilton--Jacobi 
equation \cite{MATONE}.

The quantum potential has been used in \cite{MMATONE} to derive 
the gravitational interaction. This suggests that gravitation would have a 
quantum origin. Analogous conclusions might also hold for non--gravitational 
interactions as well. This suggests that physical interactions 
do {\it not} have to be introduced by hand, through the consideration of a potential; 
rather, they {\it follow} from consistency requirements. In particular, the quantum potential 
term  that it is customary to neglect in the semiclassical limit plays a decisive 
role in determining the interaction, precisely due to the observation 
that it should {\it not} be neglected \cite{MMATONE}.

The linking of gravitational with quantum effects is of course not new.
Such was the case, {\it e.g}, in Berezin's quantisation \cite{BEREZIN}, 
where quantum numbers arose from a global metric on classical phase space.
The novelty of our approach lies in the observation that the complete 
trivialisation of the metric, {\it i.e.}, having no metric at all, can 
produce locally the same quantum numbers that follow globally from a metric 
approach such as Berezin's.

The technical aspects of our approach certainly differ from those of 
\cite{MATONE, MMATONE}. However, one can also perceive a conceptual analogy 
in the use of coordinate transformations in order to trivialise a given system.
In our context, {\it trivialisation} does not mean cancellation of the 
interaction term, as in the Hamilton--Jacobi approach. Rather, 
it refers to the choice of a vacuum around which to perform a perturbative expansion 
in powers of $\hbar$, as explained in section \ref{principle}. 

\section{Conclusions and outlook}\label{outlook}

The previous discussion motivates the following statement:

{\it Given any quantum system, there always exists a coordinate transformation 
that transforms the system into the semiclassical regime, i.e., into a system that 
can be studied by means of a perturbation series in powers of $\hbar$ 
around a certain local vacuum.}

One could turn things around and regard the previous statement as a 
starting point from which to derive a formulation of quantum mechanics such as that claimed 
by Vafa \cite{VAFA}. From this viewpoint, in this paper we have proved that the 
HFT allows for a successful implementation of the statement just formulated,
while at the same time reducing to standard quantum mechanics after the choice 
of a local vacuum. Our construction also draws attention to the fact that 
quantum corrections may depend on the observer, so that semiclassical 
expansions do not have an absolute, {\it i.e.}, coordinate--free meaning.
Looking beyond, one could perhaps regard the HFT merely as a technical tool, 
possibly not the only one, that successfully implements the above 
statement, and go on to analyse whether or not the 
previous statement may be elevated to the category of a principle. 

\acknowledgments

It is a great pleasure to thank Diego Bellisai, Kurt Lechner, Pieralberto Marchetti, 
Marco Matone, Paolo Pasti, Dmitri Sorokin and Mario Tonin for interesting discussions.
This work has been supported by a Fellowship from Istituto Nazionale di Fisica Nucleare 
(Italy).


\begin{thebibliography}{99}

\bibitem{QFTDUAL}        
L. Alvarez--Gaum\'e and F. Zamora, \hepth{9709180}.

\bibitem{STRINGDUALITY}     
J. Schwarz, \npps{55B}{1997}{1}.  

\bibitem{MTHEORY}  
A. Giveon and D. Kutasov, \rmp{71}{1999}{983}.

\bibitem{MATRIX}    
W. Taylor, \hepth{0002016}. 

\bibitem{ADS}    
O. Aharony, S. Gubser, J. Maldacena, H. Ooguri and Y. Oz, \pr{323}{2000}{183}.

\bibitem{VAFA}
C. Vafa, \hepth{9702201}.

\bibitem{MATONE}  
A. Faraggi and M. Matone,
\pla{249}{1998}{180}; 
\plb{437}{1998}{369};
\plb{445}{1998}{77};  
\plb{445}{1998}{357};
\plb{450}{1999}{34}; 
\ijmpa{15}{2000}{1869};
G. Bertoldi, A. Faraggi and M. Matone, \cqg{17}{2000}{3965}.

\bibitem{MMATONE}

M. Matone, \hepth{0005274}.

\bibitem{ALFARO}   
V. de Alfaro, S. Fubini and G. Furlan, \nc{34A}{1976}{569}.

\bibitem{RABINOVICI}   
S. Fubini and E. Rabinovici, \npb{245}{1984}{17}.

\bibitem{DFG}  
E. Deotto, G. Furlan and E. Gozzi, \hepth{9910220};
R. Britto--Pacumio, J. Michelson, A. Strominger and A. Volovich, \hepth{9911066}.

\bibitem{SOROKIN}
E.E. Donets, A. Pashnev, V.O. Rivelles, D. Sorokin and M. Tsulaia, 
\plb{484}{2000}{337}.

\bibitem{ISHAMKAKAS}
C. Isham and A. Kakas, \cqg{1}{1984}{621}; \cqg{1}{1984}{633}.

\bibitem{AFFKLAUDER}
J. Klauder, in {\it Relativity}, eds. M. Carmeli, S. Fickler and L. Witten, Plenum Press (1970).

\bibitem{1DIMAFF}
J. Klauder and E. Alaksen, \prd{2}{1970}{272};
I. Daubechies, J. Klauder and T. Paul, \jmp{28}{1987}{85}.

\bibitem{WATSON}
G. Watson and J. Klauder, \quantph{0001026}.

\bibitem{BRASIL}
I. Vancea, \plb{480}{2000}{331}; 
M. Abdalla, A. Gadelha, I. Vancea, \plb{484}{2000}{362}; 
M. De Andrade, I. Vancea, \plb{474}{2000}{46}.

\bibitem{BEREZIN} 
F. Berezin, {\it Sov. Math. Izv.} {\bf 38} (1974) 1116; {\it Sov. Math. Izv.} {\bf 39} (1975) 363;
\cmp{40}{1975}{153}; \cmp{63}{1978}{131}.

\bibitem{WITTEN}
E. Witten, \npb{311}{1988}{46}.

\bibitem{KONTSEVICH}
M. Kontsevich, \qalg{9709040};  {\it Lett. Math. Phys.} {\bf 48} (1999) 35.

\bibitem{CATTANEO}
A. Cattaneo and G. Felder, {\tt math.QA/9902090}.

\bibitem{SCHLICHENMAIER}

M. Schlichenmaier, {\tt math.QA/9910137}.

\bibitem{FELDER}
G. Felder and B. Shoikhet, {\tt math.QA/0002057}.

\bibitem{LOZANO}  
For a recent review see C. Lozano, Ph.D. thesis, \hepth{9907123}.

\bibitem{KLAUDER}   
J. Klauder and B.--S. Skagerstam, {\it Coherent States}, World Scientific, Singapore (1985);
A. Perelomov, {\it Generelized Coherent States and their Applications}, Springer Verlag Texts and
Monographs in Physics, Berlin (1986). 

\bibitem{YOSIDA} 
K. Yosida, {\it Functional Analysis}, Springer Verlag, New York (1968);
W. Rudin, {\it Real and Complex Analysis}, McGraw--Hill, London (1970), 
{\it Functional Analysis}, McGraw--Hill, New York (1973).

\bibitem{POVM}
P. Busch, P. Lahti and P. Mittelstaedt, {\it The Quantum Theory of Measurement}, Lecture Notes in
Physics m2, Springer Verlag, Berlin (1991);
P. Busch, M. Grabowski and P. Lahti, {\it Operational Quantum Physics}, Springer Verlag, New York
(1995).

\bibitem{EGUS}
I. Egusquiza and J. Muga, \pra{61}{2000}{012104} (Erratum Phys. Rev. A 61 (2000) 059901(E)). 

\bibitem{PAULI}
W. Pauli, {\it Die allgemeinen Prinzipien der Wellenmechanik}, Handbuch der Physik, ed. S.
Fl\"ugge, vol V/1, Springer Verlag, Berlin (1958).

\bibitem{GALAPON}
E. Galapon, \quantph{9908033}, \quantph{0001062}.

\bibitem{JAEKEL}
M. Jaekel and S. Reynaud, \pla{220}{1996}{10}; {\it Found. Phys.} {\bf 28} (1998) 437.

\bibitem{WIGHTMANN}
A. Wightmann, {\it Rev. Mod. Phys.} {\bf 34} (1962) 845.

\bibitem{TOLLER}
N. Pinamonti and M. Toller, \quantph{0001109}.



\end{thebibliography}
\end{document}